\documentstyle[12pt,a4wide]{article}
\newcommand\as{\alpha_S}
\newcommand\gs{g_S}
\newcommand\MeV{{\,\sl MeV}}
\newcommand\GeV{{\,\sl GeV}}
\newcommand\slv{v\kern-5pt\raise1pt\hbox{$\scriptstyle/$}\kern1pt}
\newcommand\spur{\mathop{\rm Tr}\nolimits}

\begin{document}

\thispagestyle{empty}
\begin{flushright}
MZ-TH/96-21\\
hep-ph/9609469\\
September 1996\\
\end{flushright}
\vspace{0.5cm}
\begin{center}
{\Large\bf QCD Sum Rules for Heavy Baryons}\\[.3cm]
{\Large\bf at Next-to-Leading Order in \boldmath{$\as$}}\\
\vspace{1.3cm}
S. Groote,\footnote{Groote@dipmza.physik.uni-mainz.de}
J.G. K\"orner\footnote{Koerner@dipmza.physik.uni-mainz.de}
and O.I. Yakovlev\footnote{Yakovlev@dipmza.physik.uni-mainz.de;\\\indent 
  on leave from Budker Institute of Nuclear Physics (BINP),\\\indent
  pr.\ Lavrenteva 11, Novosibirsk, 630090, Russia}\\[1cm]
Institut f\"ur Physik, Johannes-Gutenberg-Universit\"at,\\[.2cm]
Staudinger Weg 7, D-55099 Mainz, Germany\\

\vspace{1cm}

\end{center}

\begin{abstract}

We derive QCD sum rules for heavy baryons at leading order in $1/m_Q$ 
and at next-to-leading order in $\as$. The calculation involves the 
evaluation of four different perturbative three-loop diagrams which 
determine the $\as$-corrections to the Wilson coefficients  of the leading 
term in the Operator Product Expansion (OPE). From the sum rules we obtain 
estimates for the masses and the residues of the heavy baryons $\Lambda_Q$ 
and $\Sigma_Q$. The perturbative $O(\as)$ corrections to the leading order 
spectral function amount to about $100\%$, and they shift the calculated 
values for the baryon masses slightly upward. The residues are shifted 
upward by about $20-50\%$. For the bound state energy $\bar\Lambda$ given 
by the difference of the heavy baryon mass and the pole mass of the heavy 
quark $m_Q$ we obtain $m_{\Lambda_Q}-m_Q\cong 780\MeV$ and 
$m_{\Sigma_Q}-m_Q\cong 950\MeV$. For the residues we find 
$|F_\Lambda|\cong 0.028\GeV^3$ and $|F_\Sigma|\cong 0.039\GeV^3$. 

\end{abstract}


\newpage

\section{Introduction}

There has been a great deal of interest in the physics of heavy hadrons
containing one heavy quark. The Heavy Quark Effective Theory (HQET) allows
one to study the properties of the heavy hadrons in a systematic 
$1/m_Q$-expansion. The leading term of the expansion gives rise to the
spin-flavour symmetry of Heavy Quark Symmetry (HQS). The corrections to the 
leading HQS results are determined by the small expansion parameter 
$\Lambda_{\rm QCD}/m_Q$, where $\Lambda_{\rm QCD}\approx 300\MeV$ is the 
scale of low energy physics (for a review of HQET see~\cite{HQET}, for a 
review of HQS and the sum rule approach for heavy mesons see~\cite{Neubert}).

Among the well-known predictions of HQS are e.g.\ the relations between
different heavy hadron transition form factors. Take for example, the 
$\Lambda_b\to\Lambda_c$ electro-weak transitions. The six form-factors 
describing this transition are reduced to one universal Isgur-Wise function 
in the HQS limit~\cite{Georgi,Mannel,Hussein}. Even then one still remains 
with many non-perturbative parameters characterizing the process and the 
heavy baryons participating in it. These concern the functional behaviour 
of the Isgur-Wise function itself, the masses and residues of the heavy 
baryons, and, at next-to-leading order in the heavy mass expansion, the 
average kinetic and chromomagnetic energy of the heavy quark in the heavy 
baryon. All these non-perturbative parameters can be determined by using 
non-perturbative methods as e.g.\ lattice calculations, QCD sum rule 
methods~\cite{Vain} or, in a less fundamental approach, by using potential 
models. 

In the present paper we study the correlator of two heavy baryon currents 
in the HQS limit when $m_Q\to\infty$. Using the QCD sum rule method we 
calculate the masses and residues of the heavy baryons associated with
the heavy baryon currents. In its original form the QCD sum rule method 
was suggested by Shifman et al.~\cite{Vain} as a tool to investigate the 
properties of light meson systems. Later on the method was extended to the 
case of light baryons in~\cite{Ioffe,BeIo,Chung,Pivovarov}. The QCD sum rule 
approach has proven itself to be a very powerful non-perturbative QCD-based 
tool which takes into account the properties of the QCD vacuum. It allows 
one to obtain reliable estimates for hadron masses, their residues and 
their elastic as well as their transition form factors.

In the heavy-light sector the first leading order analysis (leading both in 
$1/m_Q$ as well as in $\as$) of heavy meson properties within the QCD sum 
rule approach was performed in~\cite{Shuryak}. Later on the heavy meson sum 
rule calculation was extended to include next-to-leading order radiative 
corrections. The next-to-leading order corrections proved to be rather 
important~\cite{BrGr1,Braun,Neub1}. QCD sum rules for baryons with large but 
finite masses $m_Q$ were first studied in~\cite{Block,BaDo}. Later on the 
methods of HQET were incorporated in the sum rule analysis. The leading 
order QCD sum rules for heavy baryons were first considered 
in~\cite{Shuryak,GrYa,BaDo1}, again to leading order both in $1/m_Q$ as well 
as in $\as$. Finite mass corrections to these sum rules were discussed 
in~\cite{DaHu}.

In order to improve on the accuracy of the existing QCD sum rule analysis 
of heavy baryons one needs to avail of the next-to-leading order radiative 
corrections to the sum rules. This forms the subject of the present paper. 
We calculate the QCD radiative corrections to the leading perturbative 
term in the Operator Product Expansion (OPE) and, from these, we derive 
next-to-leading order QCD sum rules for heavy baryons in the HQS limit. We 
then go on to analyze the sum rules and compute the values of the masses 
and the residues of the heavy baryons at next-to-leading order accuracy.

\newpage

The paper is organized as follows. In Sec.~2 we introduce heavy baryon 
currents as interpolating fields for the heavy ground state baryons. In 
Sec.~3 we construct the correlator of two heavy baryon currents by means of 
the OPE and define the spectral density. In Sec.~4 we present our results 
on the radiative corrections to the perturbative part of the spectral 
density and construct renormalization group invariant QCD sum rules by 
recapitulating some known results on the one- and two-loop anomalous 
dimensions of the currents. Sec.~5 contains the results of our numerical 
analysis. Sec.~6, finally, contains our summary and our conclusions. In 
Appendix A we provide a detailed collection of results on the calculation 
of the two- and three-loop contributions to the correlator of two heavy 
baryon currents. These results are quite general in that they are given for 
general space-time dimensions and for a general baryonic current structure.

\section{Baryonic currents}

The currents of the heavy baryon $\Lambda_Q$ and the heavy quark spin 
baryon doublet $\{\Sigma_Q,\Sigma_Q^*\}$ are associated with the 
spin-parity quantum numbers $j^P=0^+$ and $j^P=1^+$ for the light diquark 
system with antisymmetric and symmetric flavour structure, respectively. 
Adding the heavy quark to the light quark system, one obtains 
$j^P=\frac12^+$ for the $\Lambda_Q$ baryon and the pair of degenerate 
states $j^P=\frac12^+$ and $j^P=\frac32^+$ for the baryons $\Sigma_Q$ 
and $\Sigma_Q^*$. The general structure of the heavy baryon currents has 
the form (see e.g.~\cite{GrYa} and refs.\ therein)
\begin{equation}\label{current}
J=[q^{iT}C\Gamma\tau q^j]\Gamma'Q^k\epsilon_{ijk}.
\end{equation}
Here the index $T$ means transposition, $C$ is the charge conjugation 
matrix with the properties $C\gamma^T_\mu C^{-1}=-\gamma_\mu$ and 
$C\gamma^T_5C^{-1}=\gamma_5$, $i,j,k$ are colour indices and $\tau$ is a 
matrix in flavour space. The effective static field of the heavy quark is 
denoted by~$Q$. For each of the ground state baryons there are two 
independent interpolating currents $J_1$ and $J_2$ which both have the
appropriate quantum numbers to interpolate to the respective ground state
baryons. They are given by~\cite{Shuryak,GrYa}
\begin{eqnarray}\label{currents}
J_{\Lambda1}&=&[q^{iT}C\tau\gamma_5q^j]Q^k\varepsilon_{ijk},\qquad
J_{\Lambda2}\ =\ [q^{iT}C\tau\gamma_5\gamma_0q^j]Q^k\varepsilon_{ijk},
  \nonumber\\[7pt]
J_{\Sigma1}&=&[q^{iT}C\tau\vec\gamma q^j]
  \cdot\vec\gamma\gamma_5Q^k\varepsilon_{ijk},\qquad
J_{\Sigma2}\ =\ [q^{iT}C\tau\gamma_0\vec\gamma q^j]
  \cdot\vec\gamma\gamma_5Q^k\epsilon_{ijk},\nonumber\\
\vec J_{\Sigma^*1}&=&[q^{iT}C\tau\vec\gamma q^j]Q^k\varepsilon_{ijk}
  +\frac13\vec\gamma[q^{iT}C\tau\vec\gamma q^j]
  \cdot\vec\gamma Q^k\varepsilon_{ijk},\\
\vec J_{\Sigma^*2}&=&[q^{iT}C\tau\gamma_0\vec\gamma q^j]Q^k\varepsilon_{ijk}
  +\frac13\vec\gamma[q^{iT}C\gamma_0\vec\gamma q^j]
  \cdot\vec\gamma Q^k\varepsilon_{ijk},\nonumber
\end{eqnarray}
where $\vec J_{\Sigma^*1}$ and $\vec J_{\Sigma^*2}$ satisfy the spin-$3/2$ 
condition $\vec\gamma\vec J_{\Sigma^*i}=0$ ($i=1,2$). The flavour matrix 
$\tau$ is antisymmetric for $\Lambda_Q$ and symmetric for the heavy quark 
spin doublet $\{\Sigma_Q,\Sigma_Q^*\}$. The currents written down in 
Eq.~(\ref{currents}) are rest frame currents. The corresponding expressions 
in a general frame moving with velocity four-vector $v^\mu$ can be obtained 
by the substitutions $\gamma_0\rightarrow\slv$ and 
$\vec\gamma\rightarrow\gamma^\mu_\perp=\gamma^\mu-\slv v^\mu$. In the 
following analysis we shall be using both of these equivalent descriptions 
alternatively, i.e.\ we shall either use the static description with
$v^\mu=(1,0,0,0)$ or a moving frame description with $v^\mu=(1,\vec{v})$ 
and $\vec{v}\neq 0$.

For a general analysis it proves to be convenient to represent the general 
light-side Dirac structure of the currents in Eq.~(\ref{currents}) by an 
antisymmetrized product of $n$ Dirac matrices 
$\Gamma=\gamma^{[\mu_1}\cdots\gamma^{\mu_n]}$. 
When calculating the one- and two-loop vertex corrections one encounters 
$\gamma$-contractions of the form $\gamma_\alpha\Gamma\gamma^\alpha$ and
$\gamma_0\Gamma\gamma_0$. The $\gamma_\alpha$-contraction leads to an
$n$-dependence according to
\begin{equation}\label{Dirac1}
\gamma_\alpha\Gamma\gamma^\alpha=h\Gamma=(-1)^n(D-2n)\Gamma.
\end{equation}
The $\gamma_0$-contraction depends in addition on an additional parameter 
$s$ which takes the value ($s=+1$) and ($s=-1$) for an even or odd number 
of $\gamma_0$'s in $\Gamma$, respectively. The $\gamma_0$-contraction reads
\begin{equation}\label{Dirac2}
\gamma_0\Gamma\gamma_0=(-1)^ns\Gamma.
\end{equation}
In order to facilitate the use of Eqs.~(\ref{Dirac1}) and~(\ref{Dirac2}) we 
have compiled a table of the $(n,s)$-values relevant for the heavy baryon 
currents treated in this paper (see Table~1).
\begin{table}
\begin{tabular}{|l|c|c|l|}
\hline
$\Gamma$&$n$&$s$&particles\\\hline\hline
$\gamma^{\rm AC}_5$&$0$&$+1$&$\Lambda_1$\\\hline
$\gamma^{\rm AC}_5\gamma_0$&$1$&$-1$&$\Lambda_2$\\\hline
$\vec\gamma$&$1$&$+1$&$\Sigma_1,\Sigma^*_1$\\\hline
$\gamma_0\vec\gamma$&$2$&$-1$&$\Sigma_2,\Sigma^*_2$\\\hline\hline
$\gamma^{\rm HV}_5$&$4$&$-1$&$\Lambda_1$\\\hline
$\gamma^{\rm HV}_5\gamma_0$&$3$&$+1$&$\Lambda_2$\\\hline
\end{tabular}
\caption{Specific values of the parameter pair $(n,s)$ for particular
cases of the light-side Dirac structure $\Gamma$. $\gamma^{\rm AC}_5$ refers
to the naive $\gamma_5$-scheme with an anticommuting 
$\gamma_5$~\cite{Kreimer} and $\gamma_5^{\rm HV}$ to the $\gamma_5$-scheme 
due to Breitenlohner, Maison, 't~Hooft and Veltman~\cite{tHVt}.}
\end{table}

\section{Correlator of two baryonic currents}

In this section we describe the steps needed for the evaluation of baryonic 
QCD sum rules. One starts with the correlator of two baryonic currents,
\begin{equation}\label{correlator}
\Pi(\omega=k\cdot v)=i\int d^4xe^{ikx}\langle 0|T\{J(x),\bar J(0)\}|0\rangle,
\end{equation}
where $k_\mu$ and $p_\mu$ are the residual and full momentum of the heavy
quark and $v_\mu$ is the four-velocity using the momentum expansion   
$p_\mu=m_Qv_\mu+k_\mu$. As was mentioned before, there are two possible 
choices of interpolating currents for each of the heavy baryon states, 
given by $\Gamma_1$ and $\Gamma_2=\Gamma_1\slv$. Thus one may consider 
correlators of the same currents (diagonal correlators) or of different 
currents (non-diagonal correlators). In the general case, one may even 
consider correlators built from a linear combination $J=J_1+bJ_2$ of these 
currents with an arbitrary coefficient $b$. We mention that the choice 
$b=1$ corresponds to a constituent quark model current which has maximal 
overlap with the ground state baryons in the constituent quark model 
picture. In this paper we limit our attention to diagonal correlators only.

The correlator in Eq.~(\ref{correlator}) depends only on the energy 
variable $\omega=k\cdot v$ because of the static nature of the heavy
propagator. It can be factorized into a spinor dependent piece and a
scalar correlator function $P(\omega)$ according to
\begin{eqnarray}
\Pi(\omega)=\Gamma'\frac{1+\slv}2\bar\Gamma'
  \frac14\spur(\Gamma\bar\Gamma)2\spur(\tau\tau^\dagger)P(\omega).
\end{eqnarray}
Following the standard QCD sum rule method~\cite{Vain}, the correlator is 
calculated in the region $-\omega\approx 1-2\GeV$, including perturbative 
and non-perturbative contributions, where the non-perturbative contributions 
can in general be quite large. The non-perturbative effects are taken into 
account by employing an Operator Product Expansion (OPE) for the 
time-ordered product of currents in Eq.~(\ref{correlator}). One then has 
\begin{equation}
T\{J(x),\bar J(0)\}=\sum_d C_d(x^2)O_d,
\end{equation}
where the operators $O_d$ are local operators with a given dimension $d$, 
$O_0=\hat 1$, $O_3=\langle\bar qq\rangle$, $O_4=\langle GG\rangle$, \dots\ , 
and the expansion coefficients $C_d(x^2)$ are the corresponding coefficient
functions or Wilson coefficients of the OPE.

A straightforward dimensional analysis shows that the OPE of the diagonal
correlator contains only even-dimensional terms. We take into account the 
perturbative term for $d=0$, the gluon condensate term for $d=4$ and a 
condensate term with four quark fields for $d=6$. The four-quark operator
will be factorized into a product of two two-quark operators,
$\langle\bar q(0)q(x)\rangle^2$~\cite{Vain}. Accordingly the Fourier 
transform of the scalar correlator function $P(\omega)$ reads 
\begin{eqnarray}\label{OPE}
P(t)=P_{\rm OPE}(t)=i\theta(t)N_c!\Big(\frac1{\pi^4t^6}
  +\frac{c\as\langle GG\rangle}{32N_c(N_c-1)\pi^3t^2}
  -\frac1{4N_c^2}\langle\bar q(0)q(t)\rangle^2\Big),
\end{eqnarray}
where $c=1$ for $\Lambda_Q$, $c=-1/3$ for $\{\Sigma_Q,\Sigma_Q^*\}$
and $N_c$ is the number of colours. For the non-local quark condensate 
$\langle\bar q(0)q(t)\rangle$ one may use the OPE about 
$\langle\bar qq\rangle:=\langle\bar q(0)q(0)\rangle$, namely
\begin{eqnarray}
\langle\bar q(0)q(t)\rangle=\langle\bar qq\rangle\Big(1
  +\frac1{16}m^2_0t^2+\pi\as\langle GG\rangle\frac{t^4}{96N_c}+\ldots\ \Big),
\end{eqnarray}
where the parameter $m_0$ is defined in Eq.~(\ref{condensates}).
Alternatively one may use the Gaussian ansatz~\cite{Rady}
\begin{eqnarray}
\langle\bar q(0)q(t)\rangle
  =\langle\bar qq\rangle\exp(\frac1{16}m^2_0t^2).
\end{eqnarray}
When expanding the Gaussian ansatz one sees that the two forms agree up
to the term linear in $t^2$. Thus the two representations of the non-local
quark condensate are quite similar to one another for small values of $t$.
In our sum rule analysis we shall make use of the Gaussian ansatz because
it provides for better stability of the sum rules.

For the condensates we use the numerical values
\begin{eqnarray}
\langle\bar qq\rangle&=&-(0.23\GeV)^3,\nonumber\\
\as\langle GG\rangle&=&0.04\GeV^4,\label{condensates}\\
g_S\langle\bar q\sigma_{\mu\nu}G^{\mu\nu}q\rangle
  &=&m_0^2\langle\bar qq\rangle \quad\mbox{with\ }m^2_0=0.8\GeV^2.\nonumber
\end{eqnarray}
With these condensate values one sees that the OPE in Eq.~(\ref{OPE}) with 
Euclidian time $\tau=it$ converges nicely for $1/\tau>0.3\GeV$. In this 
region one may thus safely truncate the OPE series after the second term. 
At $1/\tau=0.3\GeV$ the contribution of the first term is two times larger 
than the last quark condensate term. Its contribution grows rapidly with 
$1/\tau$. When $1/\tau$ is further increased we see that the correlator 
becomes dominated by the perturbative contribution. For example, at 
$1/\tau=0.6\GeV$ the perturbative term is two orders of magnitude larger 
than the contribution of the condensate terms. Note, however, that at 
$1/\tau=0.4\GeV$ the contribution of the ground state to the correlator is 
ten times smaller than the contribution of the excited states and the 
continuum. This would imply that if the theoretical and phenomenological 
continuum contributions differ by about $10\%$ (and are not equal to each 
other as assumed here), this difference would induce a $100\%$ change in 
the contribution of the ground state. Thus the sum rules can only be 
trusted at values $1/\tau<0.4\GeV$ (see also the discussions of the 
numerical results of the sum rules). In the next section we will show that 
the perturbative corrections become even more important at small Euclidian 
distances in comparison to the non-perturbative condensate contributions.

As a next step one determines the spectral density using the coordinate 
space representation $P(t)$ of the current correlator. The simplest way to 
proceed is as follows. The scalar correlator function $P_{\rm OPE}(\omega)$ 
satisfies a dispersion relation
\begin{eqnarray}\label{dispersion}
P_{\rm OPE}(\omega)=P(\omega)=\int_0^\infty
  \frac{\rho(\omega')d\omega'}{\omega'-\omega-i0}+P'(\omega),
\end{eqnarray}
where $\rho(\omega)={\sl Im}(P(\omega))/\pi$ is the spectral density and
$P'(\omega)$ is a polynomial in $\omega$, which takes into account possible 
subtractions in the dispersion representation. The Fourier transform of the 
polynomial $P'(\omega)$ consists of the $\delta$-function $\delta(t)$ and 
derivatives $\delta^{(n)}(t)$ of the $\delta$-function. A comparison with 
Eq.~(\ref{OPE}) shows that one does not in fact need any subtractions. We 
therefore set $P'(\omega)=0$. Taking the Fourier transform of 
Eq.~(\ref{dispersion}) according to
\begin{equation}
P(t)=\int\frac{d\omega}{2\pi}e^{-i\omega t}P(\omega)
\end{equation}
we obtain
\begin{eqnarray}\label{dispersion1}
P(t)=i\int^{\infty}_0\rho(\omega)e^{-i\omega t}d\omega.
\end{eqnarray}
Then we analytically continue $P(t)$ from $t>0$ to imaginary times by 
introducing the Euclidian time $\tau=it$. After this transformation, 
Eq.~(\ref{dispersion1}) becomes the well known Laplace transformation. One 
may thus use an inverse Laplace transformation in order to obtain an 
Euclidean time representation of the spectral density,
\begin{eqnarray}\label{dispersion2}
\rho(\omega)
  =\frac1{2\pi}\int^{c+i\infty}_{c-i\infty}P(-i\tau)e^{\omega\tau}d\tau,
\end{eqnarray}
where $c$ is to be chosen as a real constant to the right of all 
singularities of $P(t)$. It is then easy to check that the form 
$P(t)=\theta(t)/t^n$ gives the spectral density
$\rho(\omega)=i^{n+1}\theta(\omega)\omega^{n-1}/(n-1)\!$, whereas 
$P(t)=\theta(t)t^n$ results in 
$\rho(\omega)=-(-i)^{n-1}\delta^{(n)}(\omega)$ for $n>0$. Following the 
argumentation in~\cite{BeIo} we do not include forms of the second kind 
into the spectral density $\rho$. So the leading order perturbative 
contribution and the next-to-leading order contribution of the gluon 
condensate to the spectral density are given by
\begin{eqnarray}\label{leadingSD}
\rho(\omega)&=&\rho_0(\omega)+\rho_4(\omega),\qquad\mbox{where}\\\nonumber
\rho_0(\omega)&=&\frac{\omega^5}{20\pi^4}\quad\mbox{and}\quad
\rho_4(\omega)=c\frac{\as\langle GG\rangle}{32\pi^3}\omega.
\end{eqnarray}

\section{Radiative correction to the perturbative term}

Next we consider radiative corrections to the leading order spectral density 
in Eq.~(\ref{leadingSD}). There are altogether four different three-loop
graphs that contribute to the correlator of two baryonic currents, which are
shown in Fig.~1. Contrary to the experience in the two-loop case, the most
convenient way to calculate the three-loop contributions is to evaluate
them in momentum space. The fact that all graphs in Fig.~1 have two-point 
two-loop subgraphs greatly simplifies the calculational task. One can first 
evaluate the respective subgraphs such that one remains with a one-loop 
integration. The subgraph two-loop integration can be performed by using 
the algebraic methods described in~\cite{BrGr}. It is important to note 
that the results of the two-loop integration can be expressed in terms of
a polynomial function of the external momentum that flows into the subgraph.
Hence, the remaining integration is a one-loop type integration, where the
power of one of the propagators has become a non-integer number due to the 
use of dimensional integration. The upshot of this is that all steps of
the three-loop integration can be reduced to purely algebraic manipulations.

We present the results of calculating the two-loop and three-loop 
contributions to the correlator in the form
\begin{equation}\label{results}
P(\omega)=\lambda_0C_0B_0+\lambda_1\sum_{i=1}^4C_iB_i,
\end{equation}
where we have used the abbreviations $\lambda_0=(-2\omega/\mu)^{(2D-3)}$, 
$\lambda_1=\gs^2/(4\pi)^D(-2\omega/\mu)^{(3D-7)}$, and where $D=4-2\epsilon$
is the space-time dimension. Concerning the colour structure we have defined 
the colour factors $C_i$ ($i=0,\ldots,4$) according to the labelling of 
the graphs in Fig.~1. Their values are given by $C_0=N_c!$, 
$C_1=C_2=-N_c!C_B$ and $C_3=C_4=N_c!C_F$, where $C_F=(N_c^2-1)/2N_c$ and 
$C_B=(N_c+1)/2N_c$. Values for the scalar coefficients $B_i$ defined in 
Eq.~(\ref{results}) are listed in Appendix~A.

Putting everything together, the two-loop and three-loop scalar correlation
factor $P(\omega)$ defined in Eq.~(\ref{results}) is given by
\begin{eqnarray}\label{results1}
P(\omega)\!\!&=&\!\!-\frac{32\omega^5}{(4\pi)^4}\Bigg[
  \left(\frac{-2\omega}{\mu}\right)^{-4\epsilon}
  \frac1{40}\Bigg(\frac1{\epsilon}+\frac{107}{15}\Bigg)\\&&
  +\frac{\as}{4\pi}\left(\frac{-2\omega}{\mu}\right)^{-6\epsilon}
  \Bigg(\frac{n^2-4n+6}{45\epsilon^2}
  +\frac{40\zeta(2)+61n^2-234n+396}{225\epsilon}
  +\frac{(n-2)s}{90}\nonumber\\&&
  +\frac{5(195n^2-780n+1946)\zeta(2)-2200\zeta(3)
  +4907n^2-18408n+34352}{2250}\Bigg)\Bigg].\nonumber
\end{eqnarray}
The scalar correlation function $P(\omega)$ is renormalized by the square 
of the renormalization factor $Z_J$ of the baryonic current derived
in~\cite{GrYa}. Accordingly one has
\begin{equation}
P(\omega)=Z^2_JP^{\rm ren}(\omega)\qquad\hbox{\rm with\ }
Z_J=1+\frac{\as C_B}{4\pi\epsilon}(n^2-4n+6).
\end{equation}
The multiplication of $P(\omega)$ in Eq.~(\ref{results1}) with $Z^2_J$ 
results in the cancellation of the second power in $1/\epsilon$. The 
remaining $1/\epsilon$-singularity is purely real and hence does not 
contribute to the spectral density. Since the renormalized spectral density 
$\rho^{\rm ren}(\omega)={\sl Im}(P^{\rm ren}(\omega))/\pi$ has to be 
finite, this provides a check on our calculation. The spectral density can 
be read off from Eq.~(\ref{results1}) and is given by
\begin{eqnarray}\label{ro}
\rho^{\rm ren}(\omega,\mu)&=&\rho_0(\omega)\Bigg[
  1+\frac{\as}{4\pi}r(\omega/\mu)\Bigg],\quad\mbox{where}\\
  \rho_0(\omega)&=&\frac{\omega^5}{20\pi^4}\quad\mbox{and}\quad
  r(\omega/\mu)\ =\ r_1\ln\left(\frac\mu{2\omega}\right)+r_2
  \quad\mbox{with}\nonumber\\
  r_1&:=&\frac83(n^2-4n+6)\quad\mbox{and}\quad
  r_2\ :=\ \frac8{45}(60\zeta(2)+38n^2-137n+273).\nonumber
\end{eqnarray}
The coefficient $r_1$ of the logarithmic term in Eq.~(\ref{ro}) coincides 
with twice the one-loop anomalous dimension given in Eq.~(\ref{oneloopAD}), 
as expected. The reason is that the evolution of $\rho(\omega,\mu)$ is 
controlled by the renormalization group equation and that the anomalous 
dimension of $\langle J\bar J\rangle$ and $\rho(\omega,\mu)$ coincide.

The $\as$-correction can be seen to depend on the properties of the 
light-side Dirac matrix $\Gamma$ in the heavy baryon current, as specified 
in Table~1. As an explicit result we list representations of the 
$r(\omega/\mu)$-functions of the four baryon currents in the naively 
anticommuting $\gamma_5$-scheme (AC). They read 
\begin{eqnarray}
r_{\Lambda 1}(\omega/\mu)&=&16\ln\left(\frac\mu{2\omega}\right)
+\underbrace{\frac{8(20\zeta(2)+91)}{15}}_{\approx 66.04},\nonumber\\
  \nonumber\\
r_{\Lambda 2,\Sigma 1}(\omega/\mu)&=&8\ln\left(\frac\mu{2\omega}\right)
+\underbrace{\frac{16(10\zeta(2)+29)}{15}}_{\approx 48.48},\label{ro1}\\
  \nonumber\\
r_{\Sigma 2}(\omega/\mu)&=&\frac83\ln\left(\frac\mu{2\omega}\right)
+\underbrace{\frac{8(60\zeta(2)+151)}{45}}_{\approx 44.40}. 
\nonumber
\end{eqnarray}

The results for the two different baryon currents $\Lambda_1$ and $\Lambda_2$ 
in the 't~Hooft-Veltman $\gamma_5$-scheme (HV) differ from those presented 
above. It is well known that currents in different $\gamma_5$-schemes are 
connected by a finite renormalization factor $Z$ such that
\begin{equation}\label{finiteRen}
J_{AC}=ZJ_{HV}.
\end{equation}
These finite factors also appear in the calculation of two-loop anomalous
dimensions of baryonic currents~\cite{GKY}. From the results of~\cite{GKY}
one has
\begin{equation}Z_{\Lambda 1}=1-\frac{4\as}{3\pi}\quad\mbox{and}\quad
 Z_{\Lambda 2}=1-\frac{2\as}{3\pi}.
\end{equation} 
Using these finite renormalization factors one may convert the results in 
the naively anticommuting $\gamma_5$-scheme given in Eq.~(\ref{ro1}) to the 
corresponding results in the 't~Hooft-Veltman scheme. Least the reader 
worry that we do not list the corresponding $\Sigma$-type conversion 
factors we remind him that the $\gamma_5$ in the $\Sigma$-type currents 
act on the heavy quark line and thus there are no $\gamma_5$-ambiguities.
Nevertheless, the 't~Hooft-Velman $\gamma_5$-scheme needs some counter 
terms to satisfy some kind of Ward identities. To avoid this complification, 
we will henceforth concentrate on the naively anticommuting 
$\gamma_5$-scheme, where such counterterms are not necessary at all. We 
only mention that the finite renormalization in Eq.~(\ref{finiteRen}) will 
bring the results of the two $\Gamma_5$-schemes in line.

In order to allow for a quick appraisal of the importance of the
perturbative corrections we have exhibited the numerical values of the 
second terms in Eq.~(\ref{ro1}). For $\as$ we use the running coupling 
constant, which we normalized to the value of $\as(m_Z)=0.118$ at the 
mass of the $Z$-boson for $N_f=5$ active flavours. By doing so one has 
$\as(\mu)=0.333$ at $\mu=1\GeV$ for $N_f=3$ active flavours. Using this 
value for $\as(\mu=1\GeV)$, the above results show that the perturbative 
$\as$-corrections to the spectral density amount to about $100\%$. This 
highlights the importance of perturbative QCD radiative corrections in QCD 
sum rule applications. The same observation was made in the heavy meson 
sector \cite{BrGr1,Braun,Neub1}. As in the heavy meson sector on remains 
with several unsettled questions: 

\begin{enumerate}
\item Are there any special reasons for such big QCD ``corrections''?  
\item Can we trust the QCD sum rule predictions and the $\as$-expansion\\
when the $\as$-corrections are so big?
\item How big are the $\as^2$-corrections? Is it possible to estimate them?
\end{enumerate}
These questions should be clarified in the near future.

\subsection{Residues and QCD sum rules}

To proceed with the usual QCD sum rules analysis, we evaluate the scalar 
correlator function $P(\omega)$ using the theoretical result 
$P_{\rm OPE}(t)$ given in Eq.~(\ref{OPE}) and equate this to the dispersion 
integral over the contributions of hadron states. These consist of the 
lowest lying ground state with bound state energy $\bar\Lambda$ plus the 
excited states and the continuum. To leading order in $1/m_Q$ the bound state 
energy of the ground state is defined by 
\begin{equation}
m_{\rm baryon}=m_Q+\bar\Lambda,
\end{equation}
where $m_Q$ is the pole mass of the heavy quark. Note that the leading 
order sum rules do not depend on $m_Q$ at all since the heavy mass 
dependence has been eliminated by employing the heavy mass expansion.

We assume that the continuum is given by the OPE expression above a certain 
threshold energy $E_C$~\cite{Vain}. For the hadron-side (h.s.) contribution 
to the spectral density we thus write 
\begin{equation}
\rho_{\rm h.s.}(\omega)=\rho_{\rm g.s.}(\omega)+\rho_{\rm cont}(\omega),
\end{equation}
where the contribution of the lowest-lying ground state (g.s.) baryon is 
contained in $\rho_{\rm g.s.}$ and is given by
\begin{eqnarray}
\rho_{\rm g.s.}(\omega)=\frac12F^2\delta(\omega-\bar\Lambda).
\end{eqnarray}
In this expression $F$ is the absolute value of one of the residues $F_i$ 
($i=\Lambda,\Sigma,\Sigma^*$) of the baryonic currents defined by   
\begin{equation}\label{residue}
\langle 0|J|\Lambda_Q\rangle=F_\Lambda u,\qquad
\langle 0|J|\Sigma_Q\rangle=F_\Sigma u\quad\mbox{and}\quad
\langle 0|J_\nu|\Sigma^*_Q\rangle=\frac1{\sqrt 3}F_{\Sigma^*}u_{\nu},
\end{equation}
where $u$ and $u_{\mu}$ are the usual spin-$1/2$ and spin-$3/2$ spinors.
Note that $F_{\Sigma^*}$ coincides with $F_\Sigma$ in the lowest order of 
the heavy quark mass expansion that we are working in.

As is usual we assume hadron-parton duality for the contribution of excited
states and continuum contributions and take 
$\rho_{\rm cont}(\omega)=\theta(\omega-E_C)\rho(\omega)$, where $\rho$ is 
the result of the OPE calculations given in Eqs.~(\ref{OPE}) 
and~(\ref{leadingSD}). With these assumptions we arrive at the sum rule
\begin{equation}
P_{\rm OPE}(\omega)=\frac{\frac12F^2}{\bar\Lambda-\omega-i0}
  +\int_{E_C}^\infty\frac{\rho(\omega')d\omega'}{\omega'-\omega-i0}
\end{equation}
or
\begin{equation}\label{presumrule}
\frac{\frac12F^2}{\bar\Lambda-\omega-i0}=\int_0^{E_C}
  \frac{\rho(\omega')d\omega'}{\omega'-\omega-i0}
  +P_{\rm p.c.}(\omega),
\end{equation}
where the power counting part $P_{\rm p.c.}(\omega)$ is defined as the 
Fourier transform of that part of the correlator function $P(t)$ which 
contains non-negative powers $(t^2)^n$ ($n\geq 0$). Finally we apply the 
Borel transformation 
\begin{equation}
\hat B_T=\lim\frac{\omega^n}{\Gamma(n)}\left(-\frac{d}{d\omega}\right)^n
  \qquad n,-\omega\to\infty\quad(T=-\omega/n\hbox{\rm\ fixed}) 
\end{equation}
to the sum rule in Eq.~(\ref{presumrule}). Using  
$\hat B_T(1/(\omega-\omega'))=\exp(-\omega'/T)/T$ we obtain the Borel sum 
rule
\begin{equation}\label{sumrule}
\frac12F^2(\mu)e^{-\bar\Lambda/T}
  =\int_0^{E_C}\rho(\omega',\mu)e^{-\omega'/T}d\omega'
  +\hat BP_{\rm p.c.}(T)=:K(E_C,T,\mu),
\end{equation}
where we reintroduced the $\mu$-dependence of the spectral density, which 
causes a $\mu$-dependence for the residue. The Borel-transformed 
$\hat BP_{\rm p.c.}(T)$ can be obtained directly from $P_{\rm p.c.}(t)$ by 
the substitution $t\rightarrow-i/T$ (see the discussion in~\cite{GrYa}). 
Note that the bound state energy $\bar\Lambda$ can be obtained from the 
sum rule in Eq.~(\ref{sumrule}) by taking the logarithmic derivative with 
respect to the inverse Borel parameter according to
\begin{equation}
\bar\Lambda=-\frac{d\ln(K(E_C,T,\mu))}{dT^{-1}}.
\end{equation}
Returning to the sum rule in Eq.~(\ref{sumrule}), one has
\begin{eqnarray}
\frac12F^2(\mu)e^{-\bar\Lambda/T}&=&\frac{N!}{\pi^4}\Bigg[T^6\left(
  f_5(x_C)+\frac{\as}{4\pi}\left(\left(\ln\Big(\frac\mu{2T}\Big)
  f_5(x_C)-f_5^l(x_C)\right)r_1+r_2\right)\right)\nonumber\\&&\qquad\qquad
  +cE_G^4T^2f_1(x_C)+E_Q^6\exp\left(-\frac{2E_0^2}{T^2}\right)\Bigg]
  \label{sumrulework}
\end{eqnarray}
with the polynomials $r_1$ and $r_2$ presented in Eq.~(\ref{ro}) and the
functions
\begin{eqnarray}
f_n(x)&:=&\int_0^x\frac{x'^n}{n!}e^{-x'}dx'
  \ =\ 1-e^{-x}\sum_{m=0}^n\frac{x^m}{m!},\nonumber\\
f_n^l(x)&:=&\int_0^x\frac{x'^n}{n!}\ln x'e^{-x'}dx'.
\end{eqnarray}
In order to simplify the notation we have introduced the abbreviations
\begin{equation}
x_C:=\frac{E_C}T,\quad E_0:=\frac{m_0}4,\quad
  (E_Q)^3:=-\frac{\pi^2}{2N}\langle\bar qq\rangle\quad\mbox{and}\quad
  (E_G)^4:=\frac{\pi\as\langle GG\rangle}{32N(N-1)}.\qquad
\end{equation}
The numerical analysis of the Borel sum rule is the subject of the  
section 5.

\subsection{Anomalous dimensions}

The one-loop renormalization of the effective heavy baryon currents was
considered in~\cite{GrYa}, the two-loop case was studied in~\cite{GKY}. In 
general they differ from those in conventional QCD. The one-loop anomalous 
dimension of baryonic currents, namely the first coefficient in the 
expansion $\gamma=\sum_k(\as/4\pi)^k\gamma_k$, only depends on $n$ and is 
given by~\cite{GrYa,GKY}
\begin{equation}\label{oneloopAD}
\gamma_1=-\frac43((n-2)^2+2).
\end{equation}

The general $(n,s)$-dependent formula for the two-loop anomalous dimension 
case is rather lengthy and can be found in~\cite{GKY}. As an illustration 
we list explicit values for the two-loop anomalous dimensions as calculated 
in the MS-scheme using the naive $\gamma_5$-scheme. One has (with explicit 
values given for $N_f=3$)
\begin{eqnarray}
\gamma_{\Lambda1}&=&-8\left(\frac{\as}{4\pi}\right)
  +\underbrace{\frac19(16\zeta(2)+40N_f-796)}_{\approx-72.19}
  \left(\frac{\as}{4\pi}\right)^2,\label{andimlam1}\\ 
\gamma_{\Lambda2}&=&-4\left(\frac{\as}{4\pi}\right)
  +\underbrace{\frac19(16\zeta(2)+20N_f-322)}_{\approx-26.19}
  \left(\frac{\as}{4\pi}\right)^2,\label{andimlam2}\\
\gamma_{\Sigma1}&=&-4\left(\frac{\as}{4\pi}\right)
  +\underbrace{\frac19(16\zeta(2)+20N_f-290)}_{\approx-22.63}
  \left(\frac{\as}{4\pi}\right)^2,\\
\gamma_{\Sigma2}&=&-\frac83\left(\frac{\as}{4\pi}\right)
  +\underbrace{\frac1{27}(48\zeta(2)+8N_f+324)}_{\approx 15.81}
  \left(\frac{\as}{4\pi}\right)^2.
\end{eqnarray}

\subsection{Renormalization group invariant sum rules}

It is clear that the currents $J(\mu)$ depend on the renormalization scale 
$\mu$. This dependence can be expressed by the renormalization group equation
\begin{equation}
(\mu\frac{d}{d\mu}+\gamma)J(\mu)=0,\qquad\gamma:=\frac{d\ln Z_J}{d\ln\mu},
\end{equation}
arising from the scale independence of the bare current $J_0=Z_J(\mu)J(\mu)$, 
where $\gamma$ is the anomalous dimension of the current discussed in the 
preceding subsection. To construct a renormalization group invariant 
quantity $J_{\rm inv}$, the renormalized current $J(\mu)$ is multiplied by 
some Wilson coefficient $C(\as(\mu))$, $J_{\rm inv}=J(\mu)C(\as(\mu))$, 
which is subject to the ``dual'' renormalization group equation (see 
also~\cite{Neub1,BaDo})
\begin{equation}\label{dualRGE}
(\mu\frac{d}{d\mu}-\gamma)C(\as(\mu))=0\quad\Rightarrow\quad
(\as\beta(\as)\frac\partial{\partial\as}-\gamma(\as))C(\as)=0,
\end{equation}
where $\beta:=d\ln\as/d\ln\mu=\sum_k(\as/4\pi)^k\beta_k$ is the beta 
function of QCD with
\begin{equation}\label{beta}
\beta_1=-2(11-\frac23N_f)\quad\mbox{and}\quad
\beta_2=-4(51-\frac{19}3N_f).
\end{equation}
The formal solution of Eq.~(\ref{dualRGE}) is given by
\begin{equation}
C(\as(\mu))=\exp\left(\int^{\as(\mu)}\frac{d\alpha}\alpha
  \frac{\gamma(\alpha)}{\beta(\alpha)}\right).
\end{equation}
Finally, the perturbative expansion of the beta function and the anomalous 
dimension up to second order in $\as$ gives
\begin{equation}\label{running}
C(\as(\mu))=\as(\mu)^{\gamma_1/\beta_1}\left(1+\frac{\as(\mu)}{4\pi}
  \frac{\gamma_1}{\beta_1}\left(\frac{\gamma_2}{\gamma_1}
  -\frac{\beta_2}{\beta_1}\right)\right).
\end{equation}
The first factor in Eq.~(\ref{running}) is the result of resumming the 
leading logarithmic terms $(\as\ln\mu)^n$, where the result is 
valid only in the logarithmic approximation. As Eq.~(\ref{running}) shows, 
one needs to know also the two-loop anomalous dimension of the baryon 
current in order to obtain the evolution at next-to-leading log accuracy, 
e.g.\ in the order $\as(\as\ln\mu)^n$.

The usage of the invariance property of $J_{\rm inv}$ also provides a 
connection between currents at different renormalization scales,
\begin{eqnarray}
&&J(\mu_2)C(\as(\mu_2))=J(\mu_1)C(\as(\mu_1))
  \quad\Rightarrow\nonumber\\[7pt]
  J(\mu_2)&=&J(\mu_1)C(\as(\mu_1))C(\as(\mu_2))^{-1}
  =:J(\mu_1)U(\mu_1,\mu_2),\\[7pt]
U(\mu_1,\mu_2)&=&\exp\left(\int^{\as(\mu_1)}_{\as(\mu_2)}
  \frac{d\alpha}{\alpha}\frac{\gamma(\alpha)}{\beta(\alpha)}\right)
  \nonumber\\&=&\left(\frac{\as(\mu_1)}{\as(\mu_2)}
  \right)^{\gamma_1/\beta_1}\left(1+\frac{\as(\mu_1)-\as(\mu_2)}
  {4\pi}\frac{\gamma_1}{\beta_1}\left(\frac{\gamma_2}{\gamma_1}
  -\frac{\beta_2}{\beta_1}\right)\right),
\end{eqnarray}
where $U(\mu_1,\mu_2)$ is perturbatively evaluated up to next-to-leading 
order in $\as$ (see also the discussion in~\cite{Neubert,Bardeen,JiMu}).

As being evident from Eq.~(\ref{residue}), also the residues are functions 
of the renormalization scale parameter $\mu$, the functional form for this 
dependence is the same as for the currents. So one can construct the 
renormalization group invariant $F_{\rm inv}=F(\mu)C(\as(\mu))$ by means of 
the same Wilson coefficient. A renormalization group invariant sum rule 
can then be constructed by considering the expression
\begin{equation}\label{Invsumrules}
\frac12F_{\rm inv}^2\exp(-\bar\Lambda/T)=K(E_C,T,\mu)C(\as(\mu))^2
  =:K_{\rm inv}(E_C,T).
\end{equation}
The theoretical part of the sum rule $K(E_C,T,\mu)$ depends on the 
renormalization scale $\mu$ through the QCD perturbative corrections which 
contain the logarithmic factor $\ln(\mu)$. On the other hand, the left hand 
side of Eq.~(\ref{Invsumrules}) is independent of the renormalization scale 
$\mu$ by construction, and thus the right hand side must also be 
renormalization scale independent. It is easy to check this to first order 
in $\as$ by introducing a second scale $\mu'$ and writing
\begin{equation}
\frac{\as(\mu)}{\as(\mu')}=1-\frac{\as(\mu)}{8\pi}\beta_1
  \ln(\frac{\mu'^2}{\mu^2}).
\end{equation}
Remembering that $\rho(\omega,\mu)$ in Eq.~(\ref{ro}) appears as an 
integrand of $K(E_C,T,\mu)$, one obtains cancellations (in first order of 
$\as$) of the logarithmic factors $\ln(\mu)$ in 
$\rho(\omega,\mu)C(\as(\mu))^2$ and thereby in $K(E_C,T,\mu)C(\as(\mu))^2$. 
The cancellation occurs because of $r_1=-2\gamma_1$. In this paper we will 
make no usage of the renormalization group invariant sum rule in 
Eq.~(\ref{Invsumrules}). Instead of this we analyse the sum 
rule~(\ref{sumrule}) at some fixed point $\mu'=1\GeV$ in order to estimate 
the bound state energy $\bar\Lambda$ and the residue $F(\mu')$. The value 
of the residue $F(\mu)$ at other scales can then be obtained by using the 
evolution function $U(\mu',\mu)$, while the $\mu$-independent function 
$F_{\rm inv}$ can be immediately obtained by multiplying with $C(\as(\mu))$. 

\section{Numerical results}

Let us discuss the sum rule analysis in some detail. We start by discussing
the sum rules without radiative corrections and execute the analysis in 
consecutive steps. As Eq.~(\ref{sumrulework}) shows, the analysis of the 
non-radiatively corrected sum rules does not depend on which of the two 
different current cases are being discussed. First, we analyse the 
dependence of the bound state energy $\bar\Lambda$ on the threshold 
parameter $E_C$ and the Borel parameter $T$ in a large window of parameter 
space. The aim is to try and find regions of stability in $T$ and $E_C$. 
By looking at the three-dimensional plots for $\bar\Lambda$ as functions 
of $T$ and $E_C$ we found a stability of the sum rules only in the case of 
the exponential ansatz for the non-local operator 
$\langle\bar q(0)q(x)\rangle$. Keeping in mind the rather narrow window of 
$0.3\GeV<T<0.4\GeV$ mentioned in the connection with Eq.~(\ref{OPE}) for 
the consecutive replacement $\tau\rightarrow it\rightarrow 1/T$, one ends 
up with a more reasonable discussion of the stability. We mention that the 
range of acceptable values for $T$ is extended down to $T>0.2\GeV$ when 
radiative corrections are included, which enlarge the perturbative 
contributions. This, however, does not bring in a new region of stability.
 
Returning to the analysis of the non-radiatively corrected sum rules for the 
$\Lambda_Q$-baryon, we find areas of stability around $E_C=1.2\GeV$ in the 
window $0.3\GeV<T<0.4\GeV$. The range of confidence for $E_C$ is 
$1.0\GeV<E_C<1.4\GeV$. Therefore in Fig.~2(a) we show plots for five values 
of $E_C$ around $E_C^{\rm best}=1.2\GeV$, namely for $E_C=E_C^{\rm best}$, 
$E_C=E_C^{\rm best}\pm 0.1\GeV$ and $E_C^{\rm best}\pm 0.2\GeV$. From these
curves we then can read off values for $E_C$ and $\bar\Lambda$ with good sum 
rule stability, namely
\begin{equation}\label{LambdaEnergies}
\bar\Lambda(\Lambda)=0.78\pm 0.05\GeV\qquad\hbox{\rm in the range}\qquad
  E_C(\Lambda)=1.2\pm 0.1\GeV,
\end{equation}
where the quoted errors present rough error estimates taken from Fig.~2(a) 
according to the interval for $E_C$ in Eq.~(\ref{LambdaEnergies}).
 
Next we estimate the value of the residue. The sum rules depends on the 
three parameters $\bar\Lambda$, $E_C$ and $T$. In Fig.~2(b) we plot 
$|F_\Lambda|$ for a fixed bound state energy $\bar\Lambda(\Lambda)=0.78\GeV$ 
in the indicated window for the Borel parameter $T$. The five different 
curves again correspond to the above five different values of $E_C$. Sum 
rule stability is found at
\begin{equation}
|F_\Lambda|=0.023\pm 0.002\GeV^3,
\end{equation}
where the errors again represent rough error estimates taken from Fig.~2(b).

Next we take into account the $\as$-correction to the spectral density. As 
is evident from Eq.~(\ref{ro}), the sum rule analysis now depends on which 
of the two types of baryonic currents are used. The results for the bound 
state energy for both cases are displayed in Fig.~2(c). Using the same 
analysis as for Fig.~2(a) we obtain
\begin{equation}
\bar\Lambda(\Lambda)=0.78\pm 0.05\GeV\qquad\hbox{\rm in the range}\qquad
  E_C(\Lambda)=1.1\pm 0.1\GeV.
\end{equation}
Here we note the nice technical effect that the $\as$-corrected sum 
rule is more stable at the lower value $E_C=1.1\GeV$ for the 
continuum threshold but predicts the same bound state energy~$\bar\Lambda$. 
So there occurs some ``stabilization'' in $\bar\Lambda$.

Using the central value $\bar\Lambda(\Lambda)=0.78\GeV$ one can then obtain
values for the residue looking at Fig.~2(d), which give rise to 
the value
\begin{equation}
|F_\Lambda|=0.028\pm 0.002\GeV^3.
\end{equation}

Doing the same analysis for the $\Sigma$ baryon, at leading order in $\as$
we obtain
\begin{eqnarray}
\bar\Lambda(\Sigma)&=&0.90\pm 0.05\GeV,\qquad E_C(\Sigma)=1.3\pm 0.1\GeV
  \qquad\mbox{and}\nonumber\\|F_\Sigma|&=&0.026\pm 0.002\GeV^3
\end{eqnarray}
and including the $\as$ radiative corrections we have
\begin{eqnarray}
\bar\Lambda(\Sigma)&=&0.95\pm 0.05\GeV,\qquad E_C(\Sigma)=1.3\pm 0.1\GeV
  \qquad\mbox{and}\nonumber\\ |F_\Sigma|&=&0.039\pm 0.003\GeV^3
\end{eqnarray}
The results are displayed graphically in Fig.~3(a,b) and in Fig.~3(c,d), 
respectively.

Our predictions for the bound state energy $\bar\Lambda$ combined with the 
experimental charm and bottom baryon masses may be taken to calculate the 
charm and bottom quark pole masses $m_Q$. Taking into account the 
experimental results as given by the Particle Data Group~\cite{PDG}, namely 
$m(\Lambda_c)=2284.9\pm 0.6\MeV$, $m(\Sigma^+_c)=2453.5\pm 0.9\MeV$ and 
$m(\Lambda_b)=5641\pm 50\MeV$, we obtain the pole masses 
$m_c\approx 1500\MeV$ and $m_b\approx 4860\MeV$ for the heavy quarks.
The experimental difference of $m(\Lambda_c)-m(\Sigma_c)\approx 167\MeV$ 
\cite{PDG} is quite near to our prediction 
$m(\Lambda)-m(\Sigma)\approx 170\MeV$. Here we present only central values.
As was discussed above, the accuracy of our predictions is connected with 
the internal accuracy of the QCD sum rules method (mainly because of the 
dependence on the energy of continuum) and is probably not better than 20\%.

All the results are summarized in Table~2 where we compare our results 
with the leading order results obtained in~\cite{GrYa,DaHu,Cola}. This 
concludes our analysis. 

\begin{table}
\begin{tabular}{|r||c|c|c||c|c|}\hline
  &\cite{GrYa}&\cite{DaHu}&\cite{Cola}&L.O.&N.L.O.\\\hline\hline
  $E_C(\Lambda)$&$1.20$&$1.20\pm0.15$&$1.2\pm0.1$&$1.2\pm0.1$&$1.1\pm0.1$\\
  $E_C(\Sigma)$&$1.46$&$1.30\pm0.15$&$1.4\pm0.1$&$1.3\pm0.1$&$1.3\pm0.1$\\
\hline\noalign{\smallskip}
  $\bar\Lambda (\Lambda )$&$0.78$&$0.79\pm0.05$&$0.9\pm0.1$
  &$0.78\pm0.05$&$0.78\pm0.05$\\
  $\bar\Lambda( \Sigma)$&$0.99$&$0.96\pm0.05$&
  &$0.90\pm0.05$&$0.95\pm0.05$\\
  $\bar\Lambda(\Sigma)-\bar\Lambda(\Lambda)$&$0.21$&$0.17$&&$0.12$&$0.17$\\
\hline\hline
  $|F_\Lambda|$&$2.3\pm0.5$&$1.7\pm0.6$&$2.5\pm0.5$&$2.3\pm0.1$&$2.8\pm0.2$\\
  $|F_\Sigma|$&$3.5\pm0.6$&$4.1\pm0.6$&$4.0\pm0.5$&$2.6\pm0.2$&$3.9\pm0.3$\\
\hline
\end{tabular}

\caption{Sum rule results on non-perturbative and sum rule parameters
of heavy ground state baryons. The continuum threshold parameter $E_C$,
the bound state energy $\bar\Lambda$ and the difference between the two
bound state energies are given in $\GeV$, whereas the residues are listed 
in units of $10^{-2}\GeV^3$. The value of the Borel parameter is 
$T=0.35\GeV$.}
\end{table}

\section{Conclusions}

We have considered the Operator Product Expansion of the correlator of two
static heavy baryon currents at small Euclidian distances and determined
the $\as$ radiative corrections to the first Wilson coefficient in the 
expansion. Based on this expansion we formulated and analyzed heavy baryon
sum rules for the $\Lambda$-type and $\Sigma$-type heavy baryons using
two different types of interpolating fields for the baryons in each case.
We have discussed in some detail the scale independence of the $\as$ sum
rules which requires the consideration of the anomalous dimensions of the
heavy baryon currents at the two-loop level.

Similar to the case of heavy mesons the QCD radiative correction to the
first term in the OPE is quite large and amounts to a $100\%$
change in the perturbative contribution. The radiative correction to
the perturbative term increase the calculated sum rule values for the baryon
masses by about $10\%$ and the residues by about $20-50\%$ relative to the
corresponding lowest order values. The sum rule results do not depend
very much on which of the two possible interpolating fields is used in each 
case. The sum rule analysis is, however, quite sensitive to changes in the 
assumed threshold energy of the continuum. This sensitivity is the main 
source of uncertainty in our results and is partly due to the use of 
diagonal correlators. QCD sum rules based on the diagonal correlators 
feature a leading order spectral density which grows rapidly as 
$\rho(\omega)\approx\omega^5$. This rapid growth introduces a strong 
dependence of the sum rule results on the assumed energy of continuum. 
Second, the QCD radiative correction to the leading order spectral density 
is about $100\%$ at the renormalization scale $\mu=1\GeV$. We may try to 
make the coefficient at $\as$ in these corrections to be moderate 
considering the very low renormalization scale $\mu=10\MeV$. 

We have not considered non-diagonal sum rules which come in when one 
considers correlators between two different currents with the same quantum 
numbers. These non-diagonal sum rules bring in some new features such as a 
more ``normal'' behaviour of the spectral density 
$\rho(\omega)\approx\langle\bar qq\rangle\omega^2$ and thus probably more 
moderate QCD corrections to this spectral density. On the other hand, the 
leading term for non-diagonal sum rules is proportional to the quark 
condensate, whose value $\langle\bar qq\rangle=(-0.23\pm0.02\GeV)^3$ is 
known only with an accuracy of 10\%, which gives an additional uncertainty 
in the result for non-diagonal sum rules. The analysis of the non-diagonal 
sum rules will form the subject of a subsequent paper.

\vspace{.5truecm}

\noindent{\large \bf Acknowledgments:}\smallskip\\
This work was partially supported by the BMBF, FRG, under contract 06MZ566, 
and by the Human Capital and Mobility program under contract CHRX-CT94-0579. 
We would like to thank A.~Grozin and B.~Tausk for valuable discussions.

\newpage

\section*{Appendix A: Diagrammatic contributions} 
\setcounter{equation}{0}
\def\theequation{A\arabic{equation}}

In this Appendix we collect together results on the calculation of the
two-loop and three-loop contributions to the correlator of two heavy
baryon currents. We start with the two-loop contribution depicted in
Fig.~1 ($i=0$) where one has 
\begin{eqnarray}
B_0&=&\frac{(D-2)E_2}{16(2D-7)(2D-5)(2D-3)E_1}\tilde b_0
  \spur(\bar\Gamma\slv\Gamma\slv)\qquad\mbox{with}\nonumber\\
\tilde b_0&=&\frac{E_1}{(D-4)(D-3)}.
\end{eqnarray}
We have introduced the abbreviation 
$E_n=\Gamma(1-\epsilon)^n\Gamma(1+n\epsilon)$ (with natural numbers 
$n=1,2,3,\ldots\ $) which is also used in the subsequent presentation of 
the three-loop results.

For the three-loop contributions $i=1,2$ and $4$ depicted in Fig.~1 one has
\begin{eqnarray}
B_i&=&\frac{2(D-2)(2D-7)E_3}{9(3D-11)(3D-10)(3D-8)(3D-7)E_2}
  \tilde b_i\spur(\bar\Gamma\slv\Gamma\slv)\nonumber\\
\noalign{\noindent\mbox{with}\bigskip}
\tilde b_1&=&\frac{2(D-2)E_1^2}{(D-4)^3(D-3)^2}
  -\frac{(D-2)(3D-10)E_2}{(D-4)^3(D-3)^2(2D-7)},\\
\tilde b_3&=&\frac{(D-2)E_2}{2(D-4)^2(D-3)(2D-7)},\nonumber\\
\tilde b_4&=&\frac{-(D-2)E_2}{(D-4)^2(D-3)^2(2D-7)}.\nonumber
 \end{eqnarray}
The contribution of diagram ($2$) in Fig.~1 is the most involved one. In
order to be able to write the results in a compact form we introduce the
abbreviations
\begin{eqnarray}
Q_1&=&\Gamma(1-\epsilon)^2\Gamma(1+\epsilon)/\Gamma(1-2\epsilon)
  \quad\mbox{and}\nonumber\\
Q_2&=&\Gamma(1-\epsilon)^3\Gamma(1+2\epsilon)/\Gamma(1-3\epsilon).
\end{eqnarray}
In terms of the basic structure terms
\begin{equation}
\tilde\Gamma_0=\spur(\bar\Gamma\slv\Gamma\slv),\qquad
\tilde\Gamma_1=\spur(\bar\Gamma\gamma_\mu\Gamma\gamma^\mu)
  \quad\mbox{and}\quad\tilde\Gamma_2=\spur(\bar\Gamma\gamma_\mu
    \gamma_\nu\slv\Gamma\slv\gamma^\nu\gamma^\mu),\nonumber
\end{equation}
one obtains
\begin{eqnarray}
B_2&=&\frac{E_3}{36(D-3)(3D-11)(3D-7)Q_2}
  \sum_{j=0}^2\tilde b_{2,j}\tilde\Gamma_j\qquad\mbox{with}\nonumber\\
\tilde b_{2,0}&=&\frac{12(D-2)^2Q_1^2}{(D-4)^3(D-3)^2(D-1)}
  -\frac{24D(D-2)^2Q_2}{(D-4)^3(D-1)(3D-10)(3D-8)},\nonumber\\
\tilde b_{2,1}&=&\frac{(D^2-7D+16)Q_1^2}{(D-4)^2(D-3)^2(D-1)}
  -\frac{4(D^2-4D+8)Q_2}{(D-4)^2(D-1)(3D-10)(3D-8)},\\
\tilde b_{2,2}&=&\frac{3Q_1^2}{(D-4)^2(D-3)(D-1)}
  -\frac{4Q_2}{(D-4)(D-1)(3D-10)(3D-8)}.\nonumber
\end{eqnarray}

\vspace{1cm}
\centerline{\Large\bf Figure Captions}
\vspace{.5cm}
\newcounter{fig}
\begin{list}{\bf\rm Fig.\ \arabic{fig}:}{\usecounter{fig}
\labelwidth1.6cm\leftmargin2.5cm\labelsep.4cm\itemsep0ex plus.2ex}
\item Two-loop and three-loop contributions to the correlator of two heavy 
baryon currents. (0) two-loop lowest order contribution, (1)--(4) three-loop 
$O(\as)$ contributions.
\item Sum rule results on non-perturbative parameters of the $\Lambda_Q$ as 
functions of the Borel parameter $T$. Shown are five equidistant curves 
centered around $E_C=E_C^{\rm best}$ with a distance of $100\MeV$, $E_C$ 
growing when going from bottom to top. These are in detail\\
(a) lowest order sum rule results for the bound state energy 
    $\bar\Lambda(\Lambda)$\\
(b) lowest order sum rule results for the absolute value of the residue 
    $F_\Lambda$\\
(c) $O(\as)$ sum rule results for the bound state energy 
    $\bar\Lambda(\Lambda)$\\
    for currents $J_{\Lambda1}$ (solid) and $J_{\Lambda2}$ (dashed)\\
(d) $O(\as)$ sum rule results for the absolute value of the residue 
    $F_\Lambda$\\
    for currents $J_{\Lambda1}$ (solid) and $J_{\Lambda2}$ (dashed)
\item Sum rule results on non-perturbative parameters of the $\Sigma_Q$ as 
functions of the Borel parameter $T$. Shown are five equidistant curves 
centered around $E_C=E_C^{\rm best}$ with a distance of $100\MeV$, $E_C$ 
growing when going from bottom to top. These are in detail\\
(a) lowest order sum rule results for the bound state energy 
    $\bar\Lambda(\Sigma)$\\
(b) lowest order sum rule results for the absolute value of the residue 
    $F_\Sigma$\\
(c) $O(\as)$ sum rule results for the bound state energy 
    $\bar\Lambda(\Sigma)$\\
    for the current doublets $\{J_{\Sigma1},J_{\Sigma^*1}\}$ (solid) and 
    $\{J_{\Sigma2},J_{\Sigma^*2}\}$ (dashed)\\
(d) $O(\as)$ sum rule results for the absolute value of the residue 
    $F_\Sigma$\\
    for the corrent doublets $\{J_{\Sigma1},J_{\Sigma^*1}\}$ (solid) and 
    $\{J_{\Sigma2},J_{\Sigma^*2}\}$ (dashed)
\end{list}

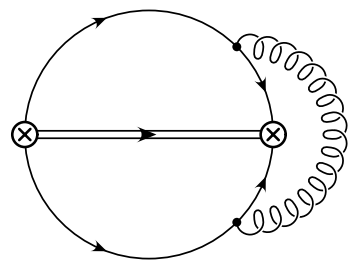
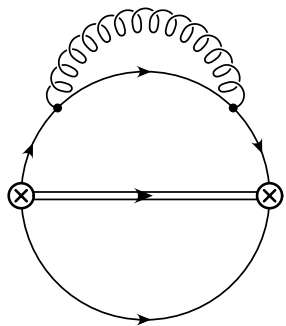
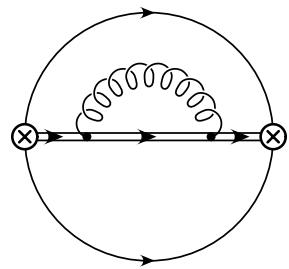
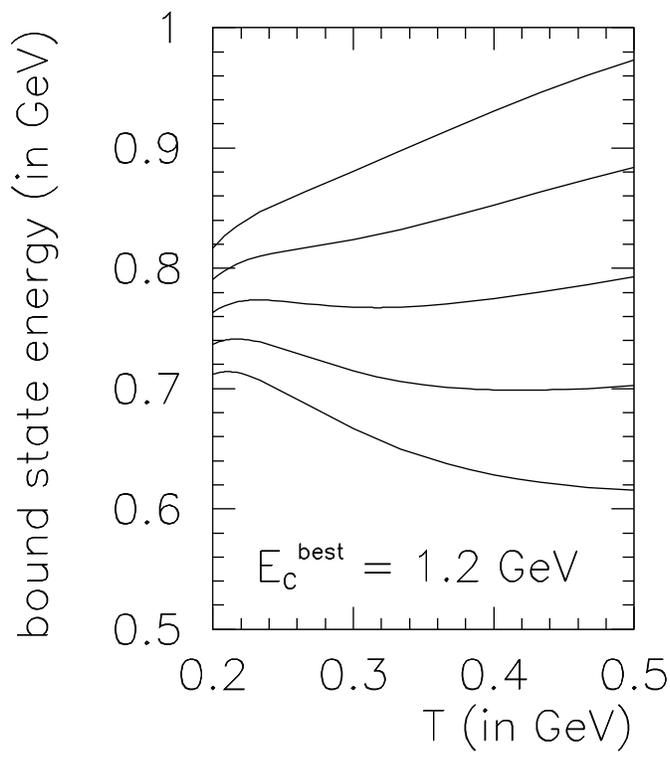
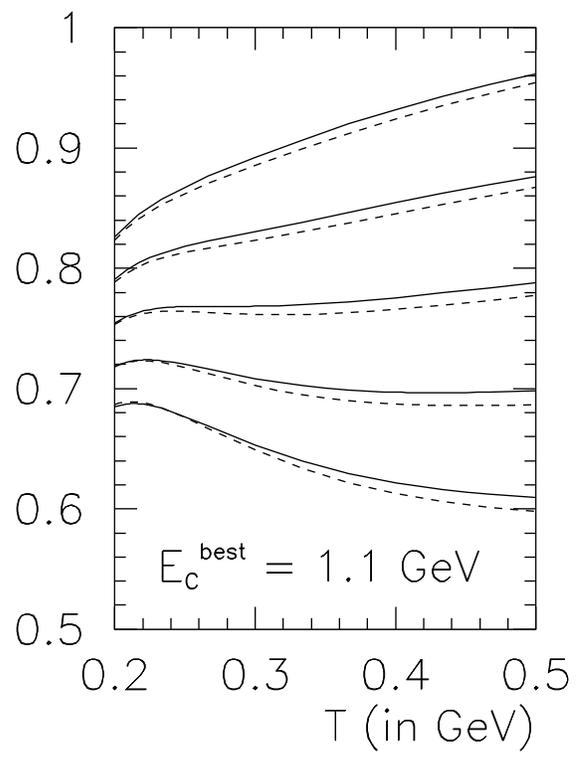
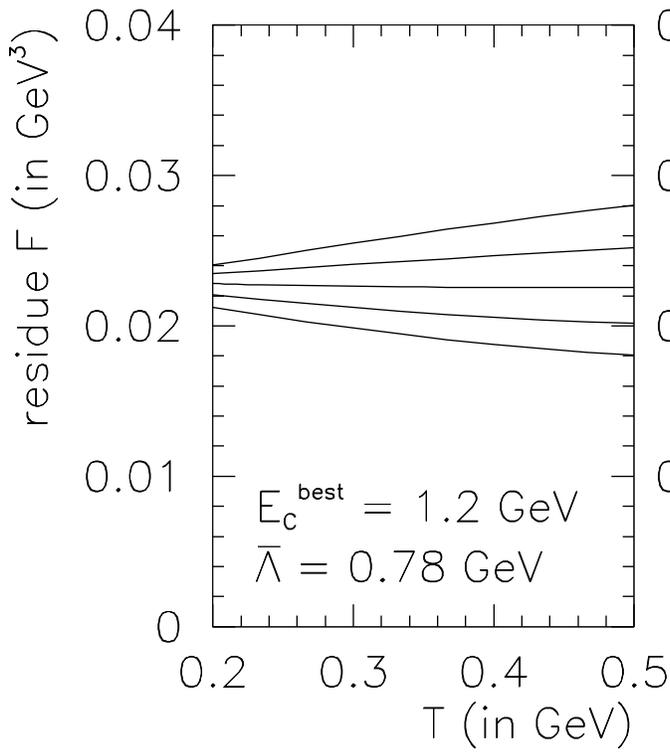
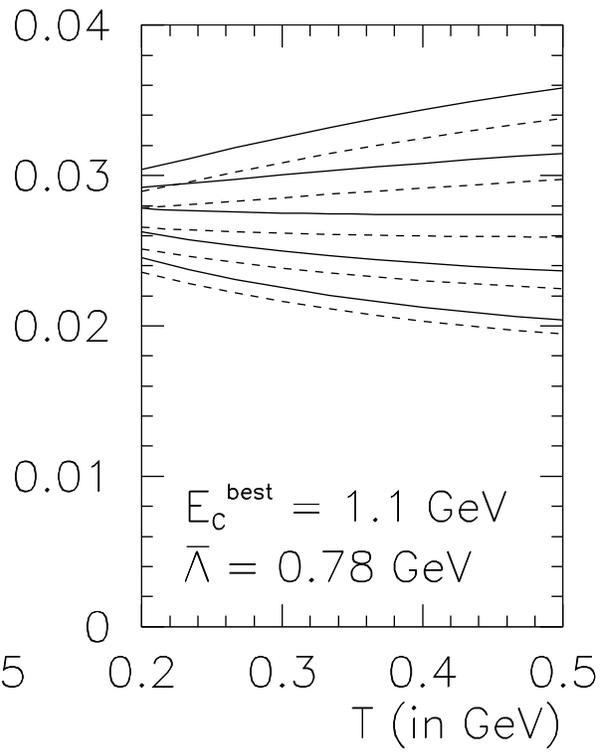
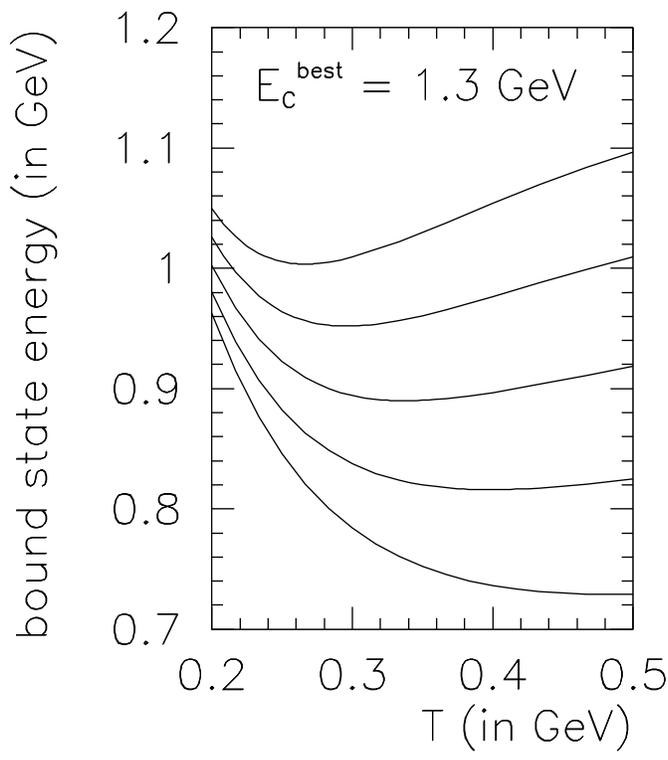
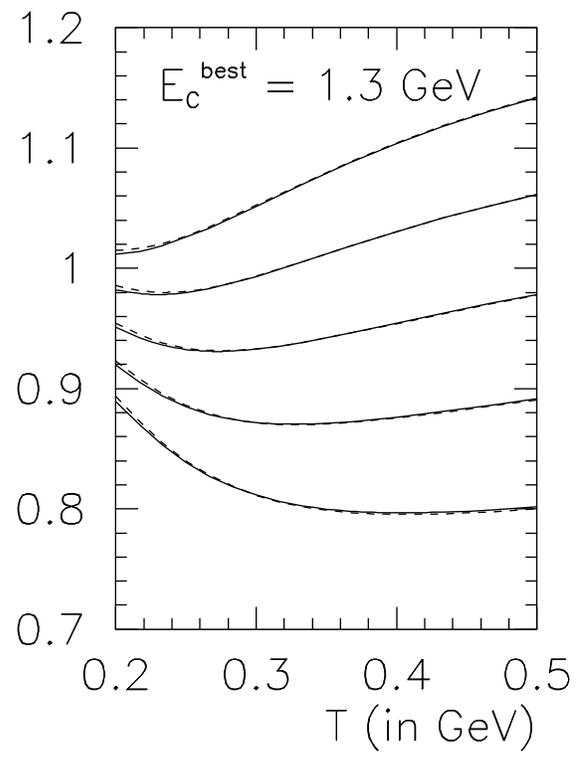
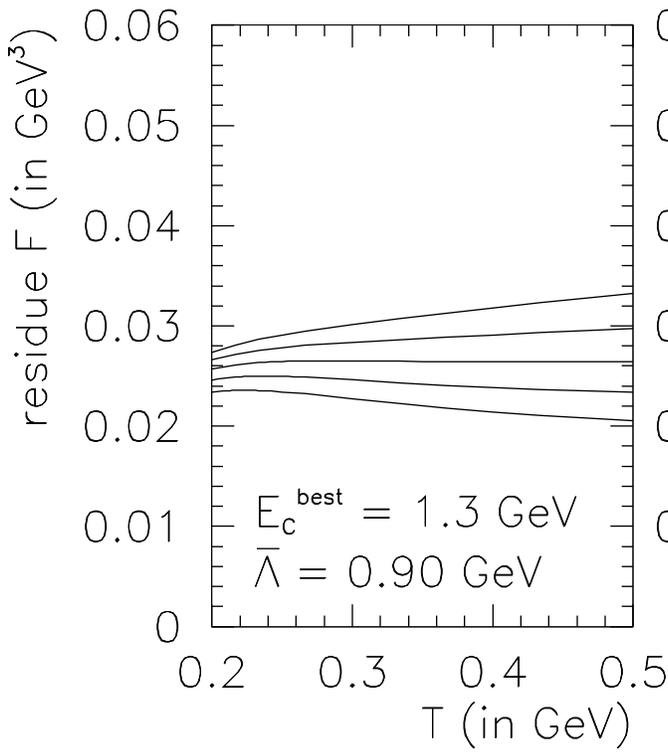
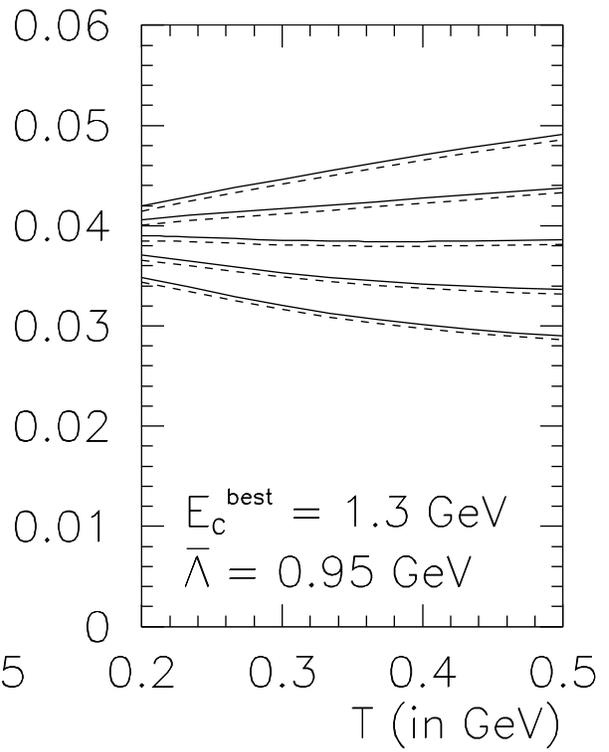
\end{document}